\documentclass[iop]{emulateapj}

\usepackage{natbib,epsfig,siunitx,url,graphicx,amsmath,footnote,float,xcolor,cases}
\usepackage[colorlinks=true,linkcolor=blue,citecolor=blue]{hyperref}

\begin{document}

\submitted{AJ; accepted}

\title{Excitation and depletion of the asteroid belt in the early instability scenario}

\author{Matthew S. Clement\altaffilmark{1,*}, Sean N. Raymond\altaffilmark{2}, \& Nathan A. Kaib\altaffilmark{1}}

\altaffiltext{1}{HL Dodge Department of Physics Astronomy, University of Oklahoma, Norman, OK 73019, USA}
\altaffiltext{2}{Laboratoire d’Astrophysique de Bordeaux, Univ. Bordeaux, CNRS, B18N, allée Geoffroy Saint-Hilaire, 33615 Pessac, France}
\altaffiltext{*}{corresponding author email: matt.clement@ou.edu}

\setcounter{footnote}{0}
\begin{abstract}

Containing only a few percent the mass of the moon, the current asteroid belt is around three to four orders of magnitude smaller that its primordial mass inferred from disk models.  Yet dynamical studies have shown that the asteroid belt could not have been depleted by more than about an order of magnitude over the past $\sim$4 Gyr.  The remainder of the mass loss must have taken place during an earlier phase of the solar system's evolution.  An orbital instability in the outer solar system occurring during the process of terrestrial planet formation can reproduce the broad characteristics of the inner solar system.  Here, we test the viability of this model within the constraints of the main belt's low present-day mass and orbital structure.  While previous studies modeled asteroids as massless test particles because of limited computing power, our work uses GPU (Graphics Processing Unit) acceleration to model a fully self-gravitating asteroid belt.  We find that depletion in the main belt is related to the giant planets' exact evolution within the orbital instability.  Simulations that produce the closest matches to the giant planets' current orbits deplete the main belt by two to three orders of magnitude.  These simulated asteroid belts are also good matches to the actual asteroid belt in terms of their radial mixing and broad orbital structure.
\break
\break
{\bf Keywords:} Asteroids, Planet Formation, Terrestrial Planets, Instabilities
\end{abstract}

\section{Introduction}

The modern asteroid belt's (AB) structure starkly contrasts that of the terrestrial and giant planet systems in that it contains less than $\sim$ 5 x $10^{-4}$ $M_{\oplus}$ of material on dynamically excited (large eccentricities and inclinations) orbits \citep{demeo13,kuchynka13}.  Though hundreds of thousands of constituents have been observed, around half of the main belt's (MB) mass is concentrated in just 4 asteroids (in order of descending mass: Ceres, Vesta, Pallas, Hygiea).  Furthermore, the composition of the belt is far from homogeneous.  The inner MB is primarily composed of silicate-rich, moderate albedo S-types, while the belt's outer regions are dominated by carbonaceous, low albedo C-types \citep{chapman75,gradie82,gradie89,bus02,demeo13}.  However, these two populations overlap substantially, and C-types account for around two thirds of all large (D $>$ 50 km), bright (H $<$ 9.7) asteroids \citep{campins18}.

While the present-day asteroid belt is well-constrained, the primordial belt is not.  In particular, there is a huge disparity in the assumed total mass in the early belt.  The primordial AB's total mass as inferred from disk models \citep{hayashi81,bitch15} was around $\sim$1-5 $M_{\oplus}$.  However, what counts from a dynamical point of view is the asteroidal mass in planetesimals and planetary embryos.  Simulations of planetesimal formation in evolving disks have found a range of outcomes (eg: \citet{carrera17} and \citet{draz18}), including the possibility of forming a ring of planetesimals around $\sim$1 AU, but none in the AB \citep{draz16}.  The low-mass AB model \citep{izidoro15,ray17sci} discussed below makes the assumption that very few planetesimals originated in the belt.

Because the primordial mass distribution is not well characterized, studying the long-term evolution of the MB can be challenging \citep{bottke15}.  Meteorites believed to originate from Vesta indicate that it differentiated and formed a crust just a few Myr \citep{shukolyukov} after calcium aluminum-rich inclusion (CAI formation; the oldest samples known in the solar system).  Thus the most massive asteroids grew to their present size when the solar system was in its infancy, and have not grown substantially larger since.  This is primarily the result of the belt's high degree of dynamical excitation greatly lengthening accretion timescales.
  
Over Gyr timescales, collisional grinding and fragmentation tend to push the belt's size distribution towards smaller diameters \citep{bottke05b}.  This seems consistent with the large number of known collisional families, particularly in the inner MB \citep{bottke06,walsh13}.  In fact, \citet{dermott18} argued that nearly all asteroids in the inner MB are members of collisional families.  Using constraints for the formation of Vesta's two enormous craters (the 505 km Rheasilvia crater on the south pole and the underlying 395 km Veneneia crater; \citet{schenk12}), \citet{bottke05a} calculated the probability that these impacts occurred during the complete collisional evolution of the MB.  If the two basins were formed in the last $\sim$2 Gyr, the probability of both events happening given the current MB size distribution would be $\sim$1$\%$ \citep{bottke15}.  The lack of other such basins on Vesta would therefore imply that the belt has always had a low mass.  Thus, from a probabilistic standpoint, it seems unlikely that the AB has lost a substantial amount of mass over the last several Gyr.  However, this implication would not hold if Vesta was implanted into the belt from a different region of the solar system \citep{bottke06_nat,mastro17,ray17}.
  
 Mean Motion (MMR) and secular resonances are the dominant mass loss mechanisms in the modern AB.  MMRs occur when an asteroid's orbital period is in integer ratio with that of another body; in this case Jupiter.  Secular resonances are the result of an object's longitude of perihelion ($\dot{\varpi}$) or longitude of ascending node ($\dot{\Omega}$) precession frequency equalling one of the solar system's dominant eigenfrequencies. Since the $\nu_{5}$ and $\nu_{6}$ secular resonances overlie several dominant MMRs with Jupiter (4:1, 3:1, 5:2, 7:3, 2:1), asteroids in these regions are quickly excited on to planet crossing orbits \citep{moons95}.  In addition to depleting the MB, these processes also form the Kirkwood gaps in the belt's orbital distribution \citep{petit01,obrien07,deienno16}.  In spite of these loss mechanisms, \citet{minton10} concluded that the MB has lost only 50$\%$ of large asteroids since attaining it's current dynamical state.  However, the simulations of \citet{minton10} only modeled large asteroids as test particles.  This means that around two to three orders of magnitude worth of AB depletion must be accounted for before the end of the planet formation epoch (for recent summaries on the evolution of the AB consult \citet{bottke15} and \citet{morby15_rev}).

Most models for terrestrial planet formation account for depletion at the 99-99.9$\%$ level with some combination of the following mechanisms: primordial depletion, giant planet influence and embryo excitation.  In the classical model of terrestrial planet formation, a population of $\sim$100 Moon to Mars massed planet-forming embryos extend throughout the inner terrestrial disk and MB region \citep{wetherill92,chambers98,chambers01,chamb_weth01,chambers07,ray09a}.  The orbital excitation provided by a population of embryos can result in substantial mass loss \citep{petit01,obrien07}.  However, these standard initial conditions fail in that they systematically produce over-massed Mars analogs and Mars to Earth massed planets in the AB \citep{chambers01,ray09a,morb12}.  Furthermore, given the strong radial dependence of embryo growth (eg: \citet{koko_ida_96,koko_ida_98,koko_ida00}), it is possible that large embryos never existed in the young belt.  Thus the most compelling solutions to the so-called ``small Mars problem'' are those that also deplete the primordial MB and sufficiently mix the radial distribution of S and C-types.  Here, we summarize the various classes of models (see \citet{ray18_rev} for a review of the models' assumptions).

$\textbf{1. Low-mass asteroid belt model:}$  \citet{iz14} showed that the inner solar system could be consistently replicated if built from a steep initial radial mass distribution.  These initial conditions, wherein the primordial Mars-forming and AB regions never contained a substantial amount of mass \citep{izidoro15}, are largely consistent with modern pebble accretion simulations of embryo and planetesimal formation \citep{levison15,draz16}.  \citet{ray17} provided the explanation for the MB's compositional dichotomy by suggesting that planetesimals near the growing giant planets could be destabilized via aerodynamic drag and scattered into the empty primordial belt.  Thus the MB's current population of C-types originated in the outer solar system prior to being implanted throughout the AB.  The major weakness of the low-mass AB model lies in the initial conditions, and whether such a steep initial mass distribution profile is realistic.

$\textbf{2. The ``Grand Tack'' hypothesis:}$  By assuming that the terrestrial planets formed out of a narrow annulus of material extending from $\sim$0.7-1.0 au, \citet{hansen09} consistently replicated the terrestrial planets' mass distribution, in particular the large mass ratios between neighboring planets (Mercury/Venus and Mars/Earth).  \citet{walsh11} provided a dynamical mechanism for these initial conditions by proposing that, during the solar system's gas disk phase, Jupiter and Saturn migrated in and out of the inner solar system \citep{masset01,morbidelli07,pierens14}.  Jupiter's presence serves to shepherd primordial MB objects on to orbits where they are scattered out of the region.  Thus, the MB was already substantially depleted during the gas disk phase.  Furthermore, \citet{deienno16} investigated the effects of the Nice Model (an orbital instability among the giant planets; the leading evolutionary model for the outer solar system \citep{gomes05,Tsi05,mor05}) on the post ``Grand Tack'' MB distribution and found that the final orbital structure was largely consistent with the current AB.  However, the high inclination orbital parameter space of the post-instability belt in that work was over-populated.  While the Grand Tack scenario succeeds at explaining the radial mixing of C and S-types, the outward migration mechanism is highly dependent on the supposed disk structure and gas accretion rates \citep{dangelo12,raymorb14,pierens14}.

$\textbf{3. An early instability:}$  The classical Nice Model \citep{gomes05,Tsi05,mor05} was originally timed in conjunction with the Late Heavy Bombardment (LHB); an inferred delayed spike in the lunar cratering record around $\sim$400 Myr after the planets formed.  Recent evidence \citep{boehnke16,zellner17,morb18,nesvorny18} has called the LHB's existence in to doubt.  Since the instability need not be tied to a specific time, moving its occurrence earlier might prevent the disruption of the fully formed terrestrial planets.  Indeed, planet ejections and collisions are common in simulations of a delayed Nice Model instability \citep{bras09,agnorlin12,bras13,kaibcham16}.  \citet{clement18} showed that timing the instability $\sim$1-10 Myr after gas disk dispersal substantially limits the mass and formation time of Mars.  This provides a natural explanation for the differences in the inferred geological formation times of Earth \citep{kleine09} and Mars \citep{Dauphas11}.  While the co-added simulations of \citet{clement18} matched the broad orbital structure of the MB, and depleted the region at greater than the $\sim$95$\%$ level, they began with unrealistic populations of large embryos and planetesimals (with the smallest simulation particles being more than an order of magnitude larger than the entire current AB mass).  In a study similar to this one, \citet{deienno18} studied the effect of an early ``Jumping Jupiter'' style instability on a primordial terrestrial disk constructed of 10,000 massless test particles.  The final MB orbital distribution of \citet{deienno18} was a good match to the real one, and the authors argued that their inferred net depletion of $\sim$90$\%$ would be consistent with the low-mass AB framework of \citet{izidoro15} and \citet{ray17sci}.  However, the \citet{deienno18} simulations only considered the ``Jumping Jupiter'' style instability \citep{bras09,nesvorny13}.  Additionally, the integrations utilized massless test particles, and rely on a primordial low mass AB to match the current low mass of the MB.

Here, we expand on the early instability framework of \citet{clement18} with detailed simulations of the scenario's consequences in the AB.  Our work differs from \citet{deienno18} in that we model the MB with fully self-gravitating bodies and include control cases with test particles. Additionally, we assume a primordial AB mass ($\sim$2 $M_{\oplus}$) consistent with the value derived from disk models.  Since each simulated instability is highly chaotic, and therefore inherently unique, our study investigates a range of instabilities.  By placing the specific dynamical state of the solar system's giant planets on our spectrum of simulated instabilities, our study seeks to infer the approximate range of MB depletion possible in an early Nice Model instability.  Thus, while \citet{deienno18} analyzed one particular instability with an outcome very similar to the modern solar system, our work scrutinizes a range of instabilities (none of which are perfect matches to the solar system) and looks for trends.

 \section{Methods}
 
We begin by selecting 8 simulations from \citet{clement18} that best replicated both the inner and outer solar system structure in accordance with the success criteria from that work.  Using the naming conventions of \citet{clement18}, these simulations are n1/1Myr/6, n2/10Myr/14, n2/10Myr/44, n2/1Myr/40, n1/10Myr/15, n1/1Myr/46, n1/10Myr/29 and n2/10Myr/25 (number of additional primordial ice giants/instability delay time/run number; henceforward referred to as runs 1-8, respectively).  We commence our study with each of these systems evolved up to the instability time.  In \citet{clement18}, instability simulations were created by first separately evolving, and then combining two independent sets of integrations.  The first was a set of giant planet resonant configurations \citep{lee02} interacting with an external disk of primordial Kuiper Belt objects \citep{nesvorny12,deienno17} up until the instability time.  The second followed the evolution of a 5 $M_{\oplus}$ terrestrial forming disk of 100 embryos and 1000 planetesimals in the presence of a static Jupiter and Saturn (locked in a 3:2 MMR).  Snapshots of the terrestrial disks at various phases of evolution (1.0 or 10.0 Myr for the systems concerned in this paper) were then added to the giant planet simulations and integrated through the Nice Model instability for an additional 200 Myr.  It is at this combination time that we select the initial conditions for the simulations of this paper.

Accurately modeling the asteroid belt with numerical integrators \citep{duncan98,chambers99,whfast} is difficult because of the vast disparity between the sizes of modern asteroids and the belt's extrapolated primordial mass.  Simulating the primordial belt using Ceres sized particles would require close to 10,000 objects; exceeding the capabilities of the conventional integrators used to study terrestrial planet formation.  Doing so using particles the size of Hygiea would involve over 100,000 individual primordial objects.  Thus, most authors are forced to model the detailed structure of the MB with massless test particles, or approximate its early state with unrealistically massive planet embryos and planetesimals.  Improving simulation resolution with test particles can be a useful tool for studying complex orbital dynamics in models where the MB is already heavily depleted in the gas disk phase due to the giant planet's influence \citep{walsh11,izidoro15,ray17,ray17sci}.  In this low surface density limit, the AB can be well characterized with test particles.  However, the early instability scenario as described in \citet{clement18} assumes no prior depletion in the AB at the start of the giant impact phase.  Because of this higher AB surface density, characterizing the levels of depletion in a fully self-gravitating AB with somewhat realistic mass resolution is an important step in validating the viability of the early instability scenario.  

To study the evolution of the MB in sufficient detail, we select the GENGA (Gravitational Encounters with GPU (Graphics Processing Unit) Acceleration) simulation package \citep{genga}.  GENGA is based on the Mercury hybrid integrator \citep{chambers99} and runs on Nvidia GPUs.  For each system, we remove all terrestrial-forming objects in the AB region (a$>$2.0 au) and replace them with 3000 primordial asteroids, each four times the current mass of Ceres.  Thus the total MB mass is $\sim$1.8 $M_{\oplus}$.  Our selection of total initial mass is derived from the average pre-instability mass of the a$>$2.0 au region in runs 1-8.

 \begin{figure}
	\centering
	\includegraphics[width=.5\textwidth]{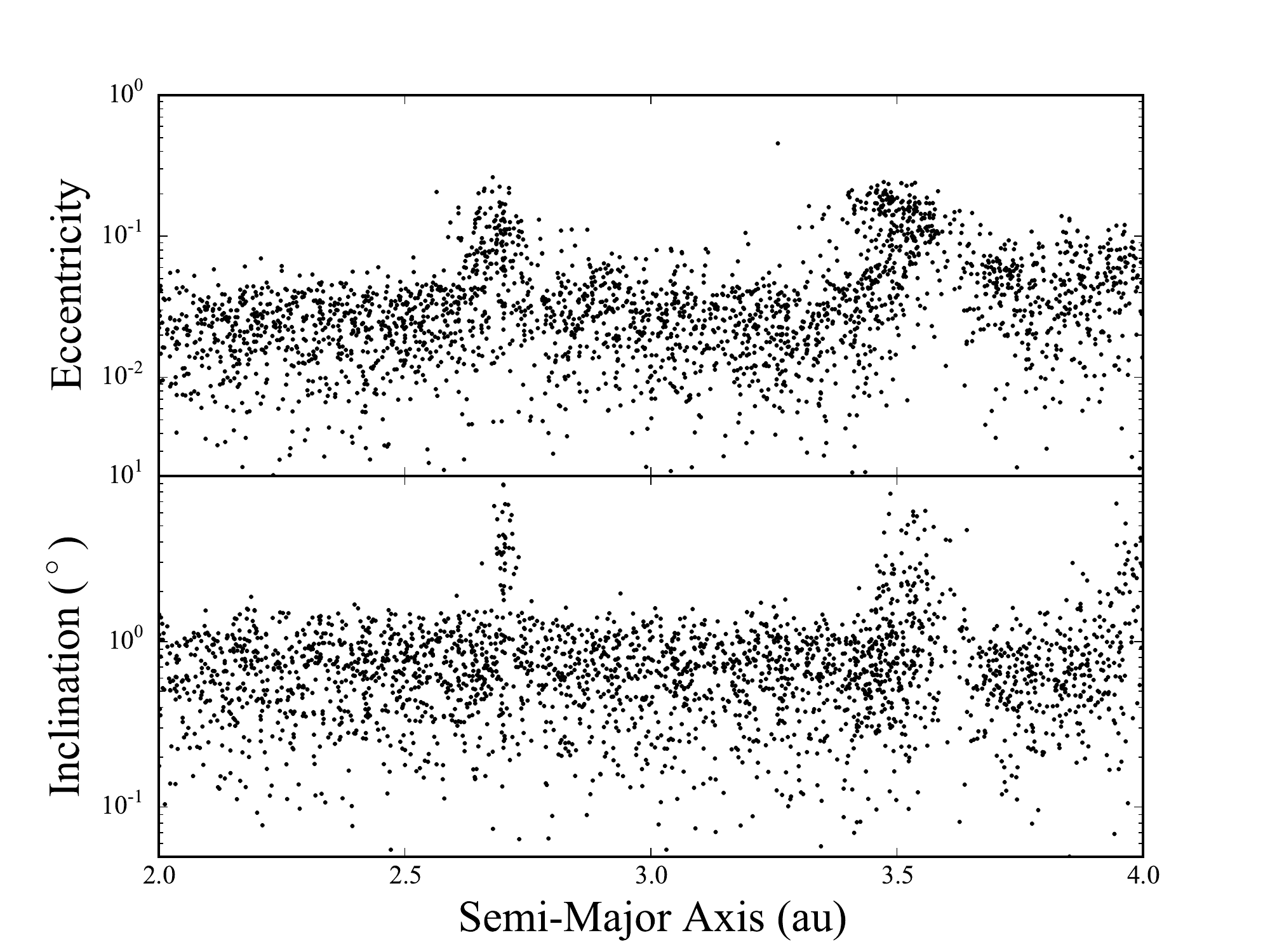}
	\caption{Pre-instability, self-stirred eccentricity (top panel) and inclination (bottom panel) distribution of our simulated AB after 1 Myr of evolution in the presence of a static Jupiter and Saturn.}
	\label{fig:stir}
\end{figure}

To create our 3000 particle ABs, we assume a surface density profile proportional to $r^{-3/2}$ \citep{birnstiel12}.  Eccentricities and inclinations are drawn from near-circular and co-planar, gaussian distributions ($\sigma_{e}=.02$ and $\sigma_{i}=.2^{\circ}$).  The remaining orbital elements are selected randomly from uniform distributions.  Next, we integrate our model AB in the presence of a static Jupiter and Saturn in a 3:2 MMR ($a_{j}$=5.6 au, $a_{s}$=7.6 au) for 1.0 Myr using a 50 day time-step.  305 accretion events occur during this initial phase of evolution, with the largest object growing to five times its initial size.  These lengthy accretion times are roughly consistent with the results of semi-analytical studies of runaway growth and embryo accretion \citep{koko_ida_96,koko_ida_98}.  However, we do note that such oversimplified initial conditions might influence our results.  Indeed, constructing the primordial belt strictly out of large bodies might overestimate the effects of planetesimal scattering and dynamical spreading in the disk.  Furthermore, our simulations are not designed to simulate fragmenting and hit-and-run collisions.  Because we are studying MB dynamics in the high surface density limit, not accounting for the dynamical damping effects of such collisions \citep{chambers13} might artificially increase the orbital excitation in our fully evolved systems.

The eccentricity and inclination structure of our self-stirred initial MB is plotted in figure \ref{fig:stir}.  We perform 18 separate, 200 Myr integrations (table \ref{table:ics}) of this AB imbedded within our respective Nice Model instability simulations using a 6 day time-step as follows (note that each individual simulation contains over 4000 objects including the forming inner terrestrial disk, 3000 asteroids, unstable giant planets, and primordial Kuiper Belt):

\begin{table*}
\centering
\begin{tabular}{c c c c}
\hline
Name & Runs & 0.0006 $M_{\oplus}$ asteroids & 0.055 $M_{\oplus}$ embryos\\
\hline
Interact & 1-8 & 3000 self-gravitating & 0  \\
Test Particle & 1a-8a & 3000 test particles & 0 \\
Embryo & 1b,2b & 3000 self-gravitating & 4\\
\hline
\end{tabular}
\caption{Summary of our three different simulation batches. The columns are as follows; (1) the name of the set of simulations, (2) the numbering scheme for the set's runs, (3) the number of 0.006 $M_{\oplus}$ asteroids, and (4) the number of 0.055 $M_{\oplus}$ embryos.}
\label{table:ics}
\end{table*}

1) 8 systems (runs 1-8) are integrated with all simulation objects treated as fully self-gravitating.

2) The same 8 systems (now denoted runs 1a-8a) are integrated with the asteroids treated as test particles, and all other objects as fully self-gravitating.  Simulations presented in \citet{levison11} indicated that disk self gravity is important in driving the evolution of the young Kuiper Belt.  By performing two sets of simulations both with and without asteroid self-gravity, we are able to test whether accounting for self-gravity affects the overall depletion or orbital structure in the MB.

3) For systems 1 and 2, we perform an additional batch of integrations (runs 1b and 2b) where we embed 4 Mercury massed planet embryos within the MB.  This allows us to probe whether embedded embryos are necessary for the success of the early instability scenario (\citet{clement18} found that much of the depletion in the scenario was due to embryo excitation).

\section{Results and Discussion}

Traditionally, the success criteria considered by dynamical studies of the AB's early evolution fall in to three general categories: depletion, radial mixing of C and S-types, and orbital structure (specifically producing a dynamically excited MB without overpopulating the high inclination parameter space above the $\nu_{6}$ secular resonance).  In the subsequent three sections, we address each of these constraints individually.

\subsection{Depletion}

 \begin{figure}
	\centering
	\includegraphics[width=.5\textwidth]{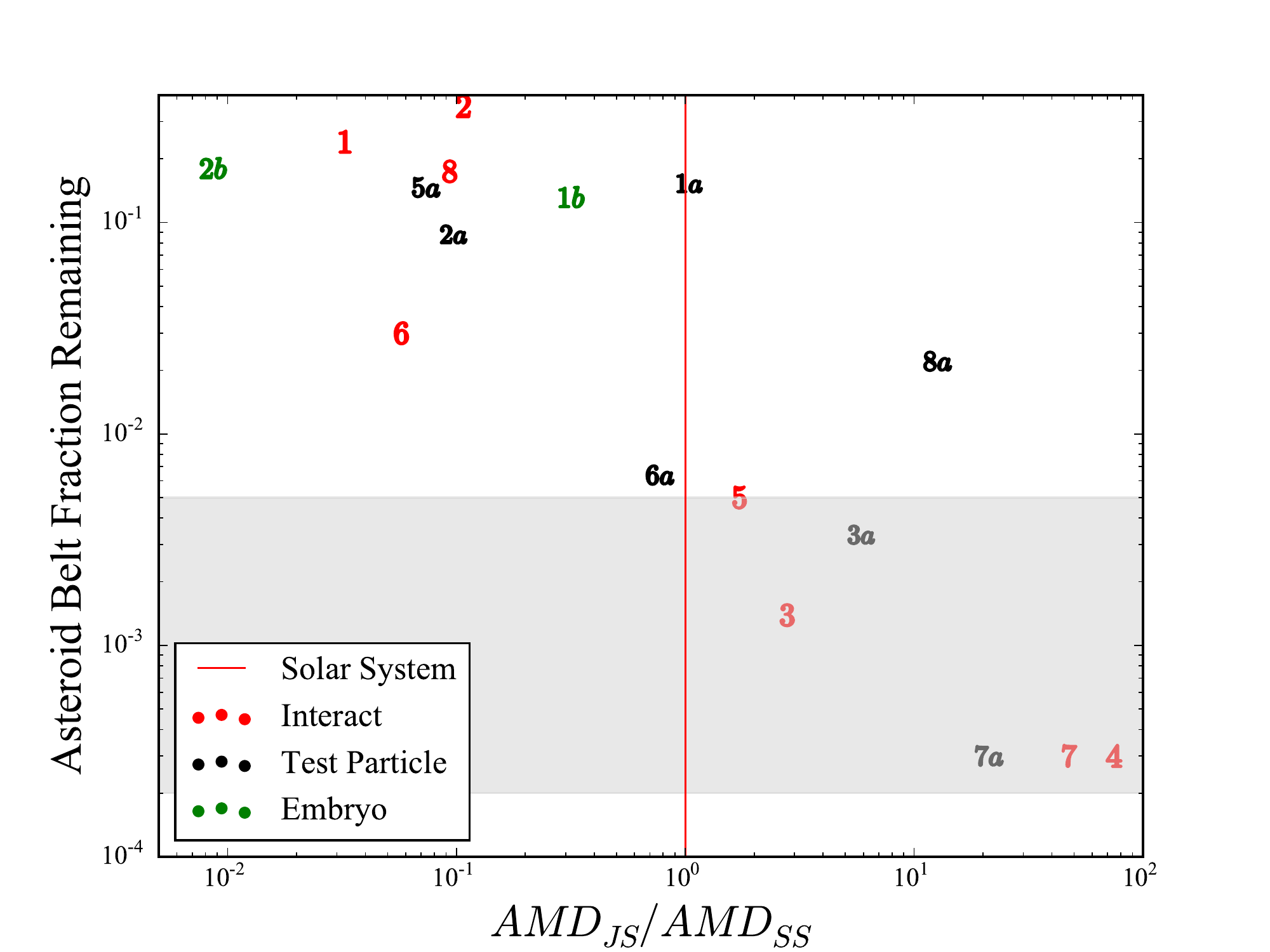}
	\qquad
	\includegraphics[width=.5\textwidth]{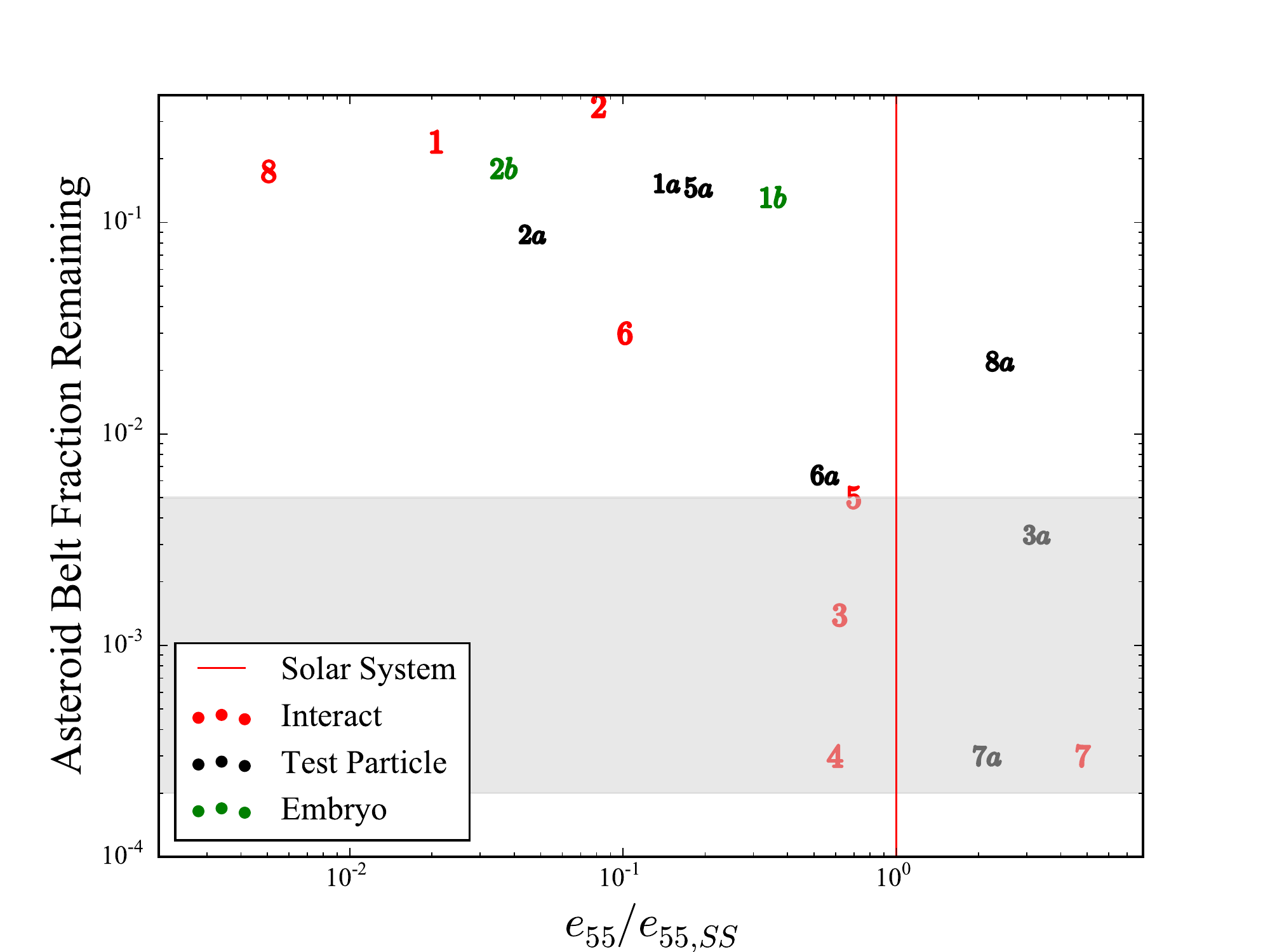}
	\qquad
	\includegraphics[width=.5\textwidth]{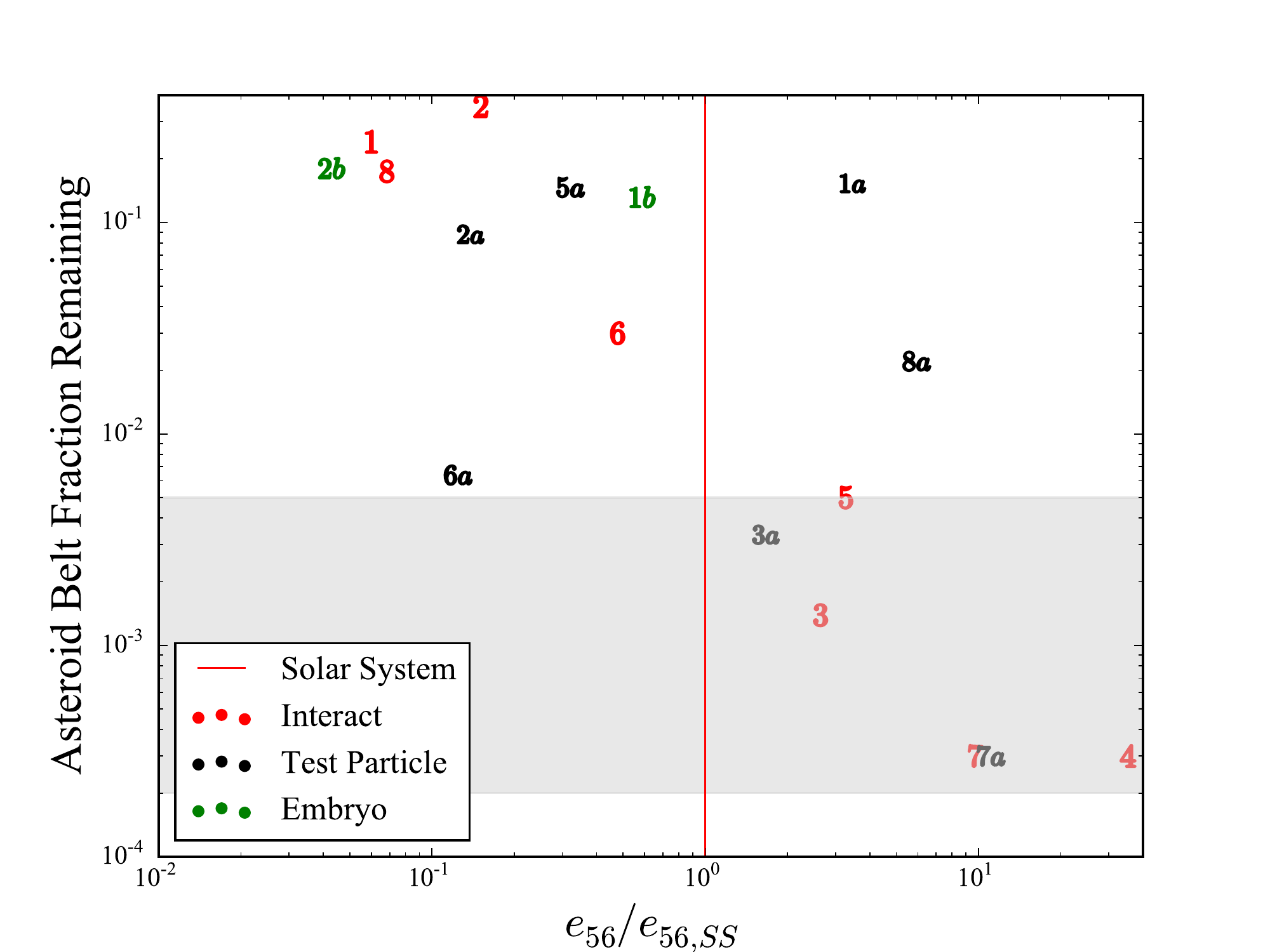}
	\caption{Dependence of AB depletion on the AMD of Jupiter and Saturn (top panel), the excitation of Jupiter's $g_{5}$ mode ($e_{55}$, middle panel) and Saturn's forced eccentricity ($e_{56}$, bottom panel).  Runs 1-8 (fully self gravitating AB) are plotted in red.  Runs 1a-8a (test particle AB) are plotted with black.  Runs 1b and 2b (embedded planet embryos) are plotted in green.  The red line denotes the solar system value for each statistic and the grey region denotes the area within a factor of 5 of a $10^{-3}$ depletion factor.  Note, run 4a is not plotted since Saturn was ejected.}
	\label{fig:e55}
\end{figure}

Since each instability is inherently unique, it is important to place the solar system outcome on the spectrum of instabilities our systems experience (and corresponding levels of depletion).  It is important to note here that none of our final giant planet configurations simultaneously replicate all the important qualities of the modern outer solar system (eg: success criteria A-D in \citet{nesvorny12} and E in \citet{deienno17}).  Of our 18 simulations, only 7 finish with the correct number of giant planets.  One simulation finished with 5 giant planets, and the instability of another (run 4a) was so violent that Saturn was ejected.  However, as shown directly in \citet{deienno18}, the perturbative effects of the ice giants on the MB are relatively weak compared to that of Jupiter and Saturn \citep{morb09}.  Therefore, systems with significantly different ice giant evolutionary schemes might still be good analogs to the solar system if their Jupiter/Saturn systems are similar.  The ratio of Saturn's orbital period to that of Jupiter is $\sim$2.48, and 8 of our systems have period ratios between 1.95-2.65 (note, $N_{GP}=$4 and $P_{S}/P_{J}<$2.8 are success criteria A and D in \citet{nesvorny12} and \citet{deienno17}, respectively).

To compare the final orbital structure of our various Jupiter/Saturn systems with the solar system (besides run 4a that lost Saturn), we begin by calculating the normalized Angular Momentum Deficit (AMD, equation \ref{eqn:amd}) for each system \citep{laskar97}.  AMD quantifies the degree to which a system of orbits differs from that of a co-planar, circular system.  Thus a simulation that finishes with an $AMD_{JS}$ close to that of the modern solar system would possess a Jupiter/Saturn pair with combined dynamical excitation similar to the actual planets.  We plot the relationship between AB depletion and $AMD_{JS}$ in the top panel of figure \ref{fig:e55}.  There is significant scatter in the data because the instability is a highly chaotic event.  However, the overall trend is very clear.  Systems that finish with a more dynamically excited Jupiter/Saturn pair also experience greater depletion in the AB.  Given these results, the systems that best match the solar system's value for $AMD_{JS}$ experience one to three orders of magnitude worth of depletion in the AB.

\begin{equation}
	AMD = \frac{\sum_{i}m_{i}\sqrt{a_{i}}[1 - \sqrt{(1 - e_{i}^2)}\cos{i_{i}}]} {\sum_{i}m_{i}\sqrt{a_{i}}} 
	\label{eqn:amd}
\end{equation}

\begin{figure}
	\centering
	\includegraphics[width=.5\textwidth]{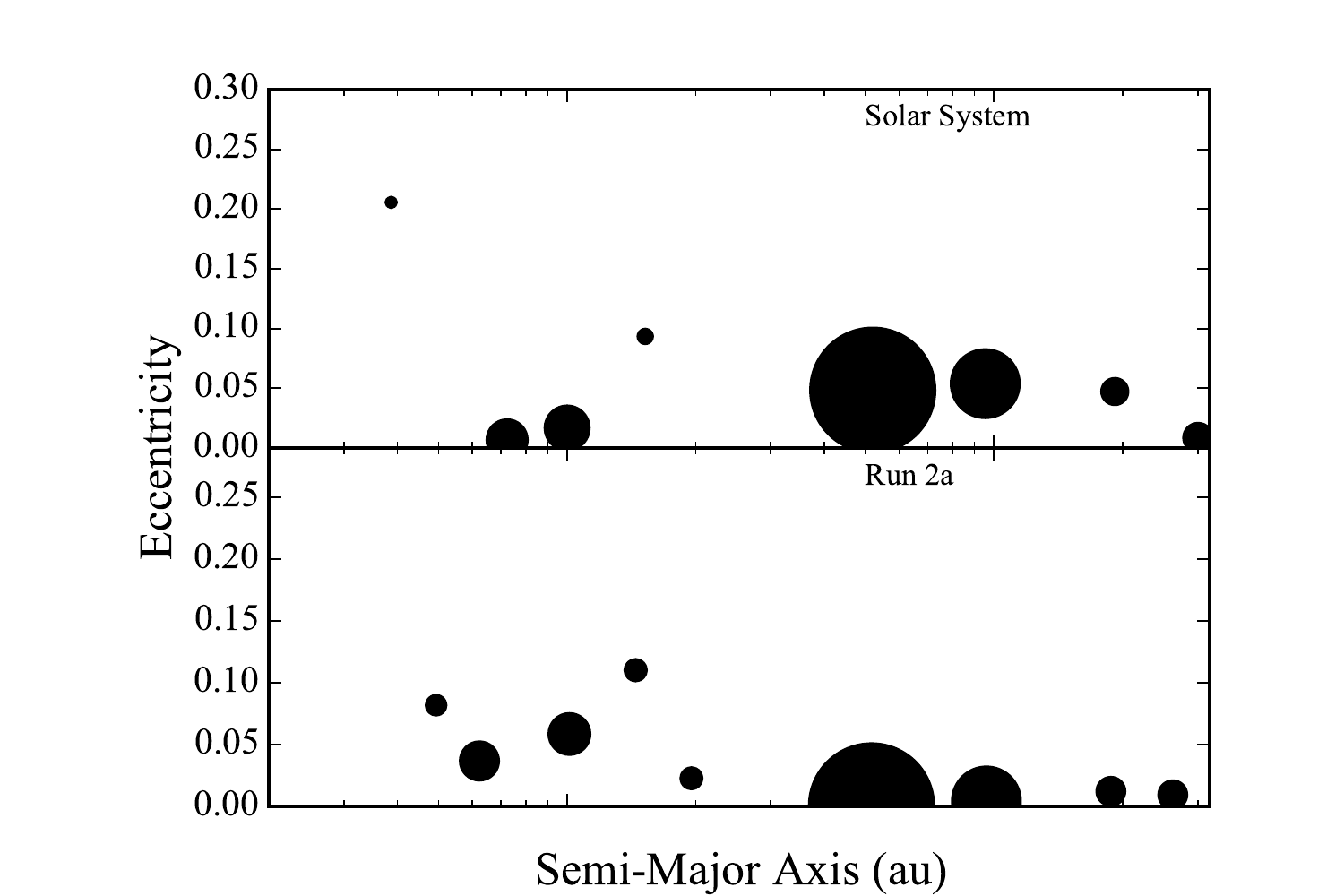}
	\caption{Semi-Major Axis/Eccentricity plot depicting the final planets from run 2a (bottom panel), compared with the actual solar system (top panel).  The size of each point corresponds to the mass of the particle (because Jupiter and Saturn are hundreds of times more massive than the terrestrial planets, we use separate mass scales for the inner and outer planets).  The final terrestrial planet masses are 0.20, 0.75, 0.86, 0.23 and 0.23 $M_{\oplus}$ respectively.}
	\label{fig:comp}
\end{figure}

However, many of our systems with values of $AMD_{JS}$ similar to that of the actual solar system are poor analogs for other reasons.  In particular, the secular architecture of the solar system is dominated by the highly excited $g_{5}$ mode (success criteria C in \citet{nesvorny12}).  For a complete discussion of the solar system's secular structure consult \citet{bras09} and \citet{morb09}.  To compare the secular landscapes of our simulated systems with the solar system we calculate $e_{55}$ (the amplitude of the Jupiter's $g_{5}$ eigenmode; $e_{55,SS}=$0.044) and $e_{56}$ (the amplitude of Saturn's forcing on Jupiter's eccentricity; $e_{56}=$0.016) for each integration via Fourier analysis \citep{nesvorny96}.  The bottom two panels of figure \ref{fig:e55} show the dependence of AB depletion on the values of $e_{55}$ and $e_{56}$.  Again, in spite of significant data scatter and small number statistics, both plots show the same general trend of increased secular amplitudes correlating with larger AB depletion fractions.  The relationship between $e_{55}$ and AB depletion is perhaps the clearest of the three panels in figure \ref{fig:e55}.  Multiple previous studies have concluded that the sufficient excitation of $e_{55}$ is supremely important in driving the secular evolution of the solar system \citep{morb09,nesvorny12,clement18}.  Therefore, it is promising that all but one of the systems in our study that sufficiently excite $e_{55}$ to within a factor of two of its present value experience at least two orders of magnitude worth of depletion in the AB.  This depletion, which is related to orbital excitation, is thus largely driven by $e_{55}$ transmitting excitation to the MB via large forced vectors during the instability phase.  Our results therefore indicate that the early instability scenario should be able to account for AB depletion at the 99-99.9$\%$ level.  As an example, run 3 finishes with $N_{GP}=$4, $P_{S}/P_{J}=$2.3, $AMD_{JS}=$2.7 x $AMD_{JS,SS}$, $e_{55}=$0.62 x $e_{55,SS}$ and 0.13$\%$ of AB objects remaining.  Furthermore, achieving this level of AB depletion and successfully forming the inner solar system are not mutually exclusive traits in our simulations.  To illustrate this, figure \ref{fig:comp} plots the final planets from run 2a, compared with the current solar system.  The stochastic nature of planet formation makes it impossible to select a perfect solar system analog out of just 18 simulations.  In run 2a the eccentricities of Earth and Venus are too large, the first three outer planets are under-excited, Mercury is over-massed, Neptune's semi-major axis is incorrect, and an extra Mars analog forms near the inner edge of the AB.  However, we still plot this result to demonstrate how the run depletes 91.2$\%$ of the AB's mass without totally inhibiting the formation of the terrestrial planets.

It is also important to point how challenging it is for N-body simulations of the Nice Model instability to simultaneously replicate all facets of the outer solar system \citep{nesvorny12,gomes17,deienno17}.  In particular, few simulations are able to sufficiently excite $AMD_{JS}$ and $e_{55}$ while keeping $P_{S}/P_{J}$ low.  \citet{nesvorny12} performed over 5,000 instability simulations using a variety of different giant planet configurations and only 5$\%$ of the most successful batch simultaneously met constraints for $e_{55}$ and $P_{S}/P_{J}$ (success criterion C and D in that work).  Within our sample of 18 instabilities, only one (run 3) successfully excites the orbits of Jupiter and Saturn while maintaining $P_{S}/P_{J}<$2.8.  Though it is encouraging that this run depletes the MB at close to the 99.9$\%$ level, our results should still be taken in context with the fact that all our other runs that experience three orders of magnitude worth of MB depletion (table \ref{table:nu6}) finish with $P_{S}/P_{J}>$2.8.

Though the differences in depletion trends between our fully self-gravitating and test particle ABs are not statistically significant, the null result is still important.  Nearly all dynamical studies of early depletion in the AB only consider test particles \citep{obrien07,deienno16,izidoro16,deienno18}.  As discussed in the section 2, test particles provide a good approximation in the low surface density limit.  However, since we assume no prior depletion in the AB, including the effects of asteroid self-gravity could be important when studying the total early depletion.

\subsection{Radial Mixing}

It is inherently difficult for N-body studies of terrestrial planet formation to explain the compositional dichotomy between S and C-types because there is no consensus as to where each type originated from in the disk.  However, nucleosynthetic differences between ordinary and carbonaceous chondrite groups \citep{burkhardt11} indicate that they formed at different radial distances \citep{kruijer17}.  Early studies of planet formation in the inner solar system assumed that the different populations exist today because of a primordial division (commonly referred to as a``snow line'') in the material \citep{wetherill92,chambers98}.  \citet{grimm93} proposed that S-types formed inside $\sim$2.7 au because of the higher probability of capturing energy from radionuclides (eg: $^{26}$Al).  In the same manner, C-types formed exterior to $\sim$2.7 au where $^{26}$Al is extinct.  However, whether ice exists at a snow line depends on the drift of small particles \citep{ciesla06}.  Studies of disk dynamics seem to indicate that snow lines tend to migrate inward \citep{lecar06,kennedy08,martin12}.  Further complicating the problem, this migration can be blocked by the growing massive planets \citep{lambrechts14a}, and thus the condensation temperature at any given point in the disk may not be correlated with the presence of volatiles \citep{morby16_ice}.  Alternatively, \citet{ray17} proposed that the AB could have already been populated with material from the outer disk ($\sim$5-20 au) via scattering events during the giant planet growth phase.

 \begin{figure*}
	\centering
	\includegraphics[width=.49\textwidth]{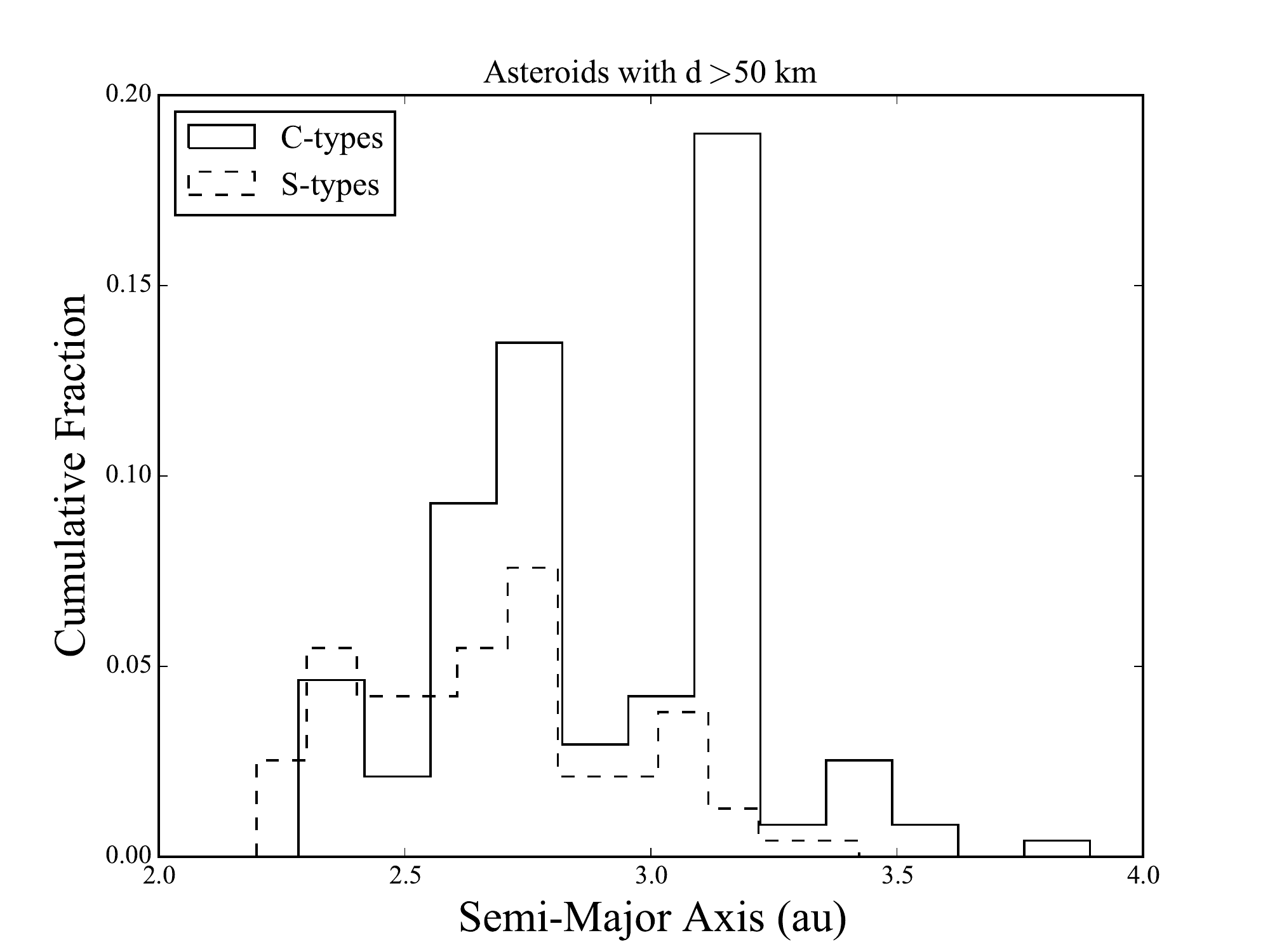}
	\includegraphics[width=.49\textwidth]{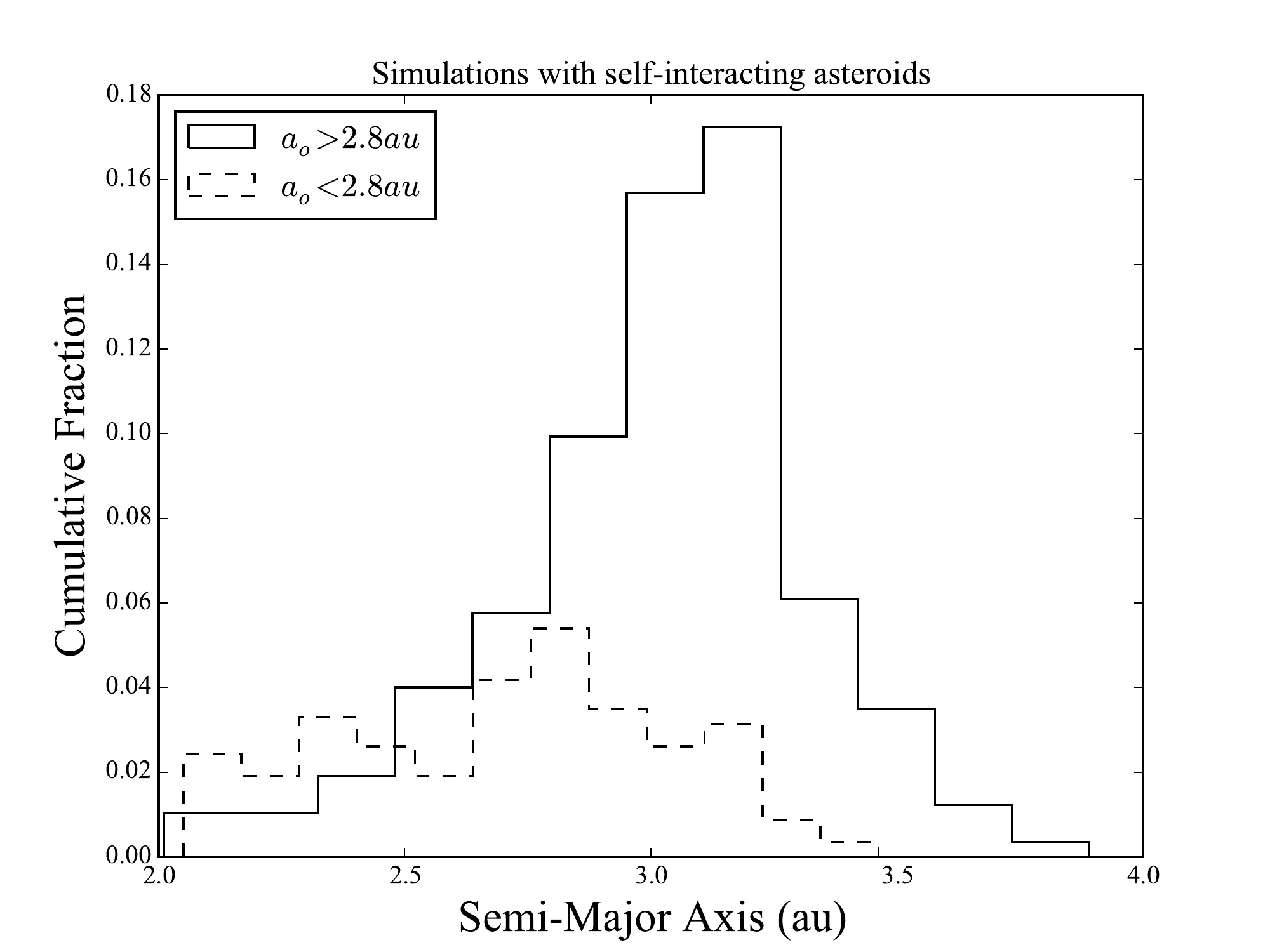}
	\qquad
	\includegraphics[width=.49\textwidth]{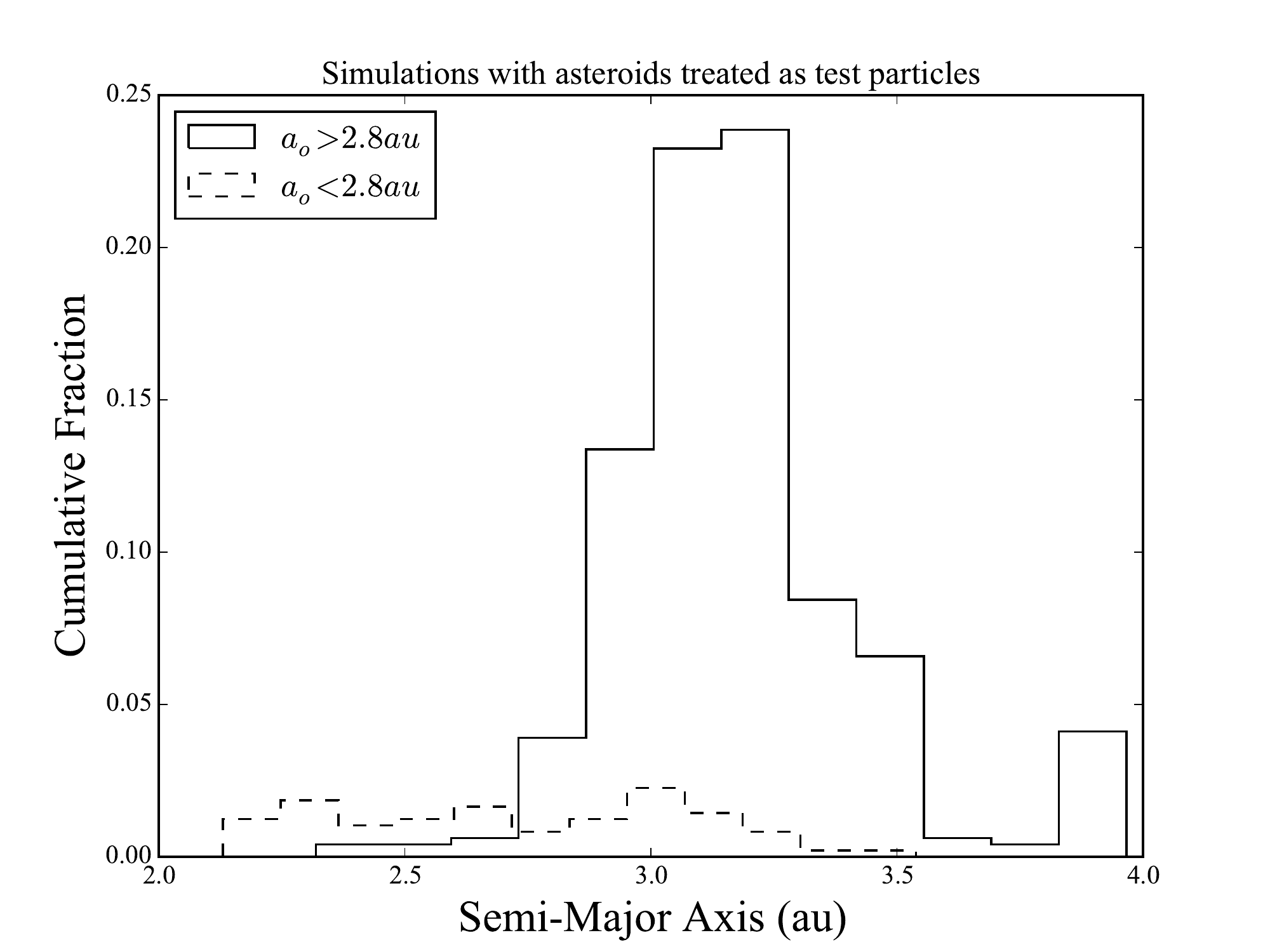}
	\includegraphics[width=.49\textwidth]{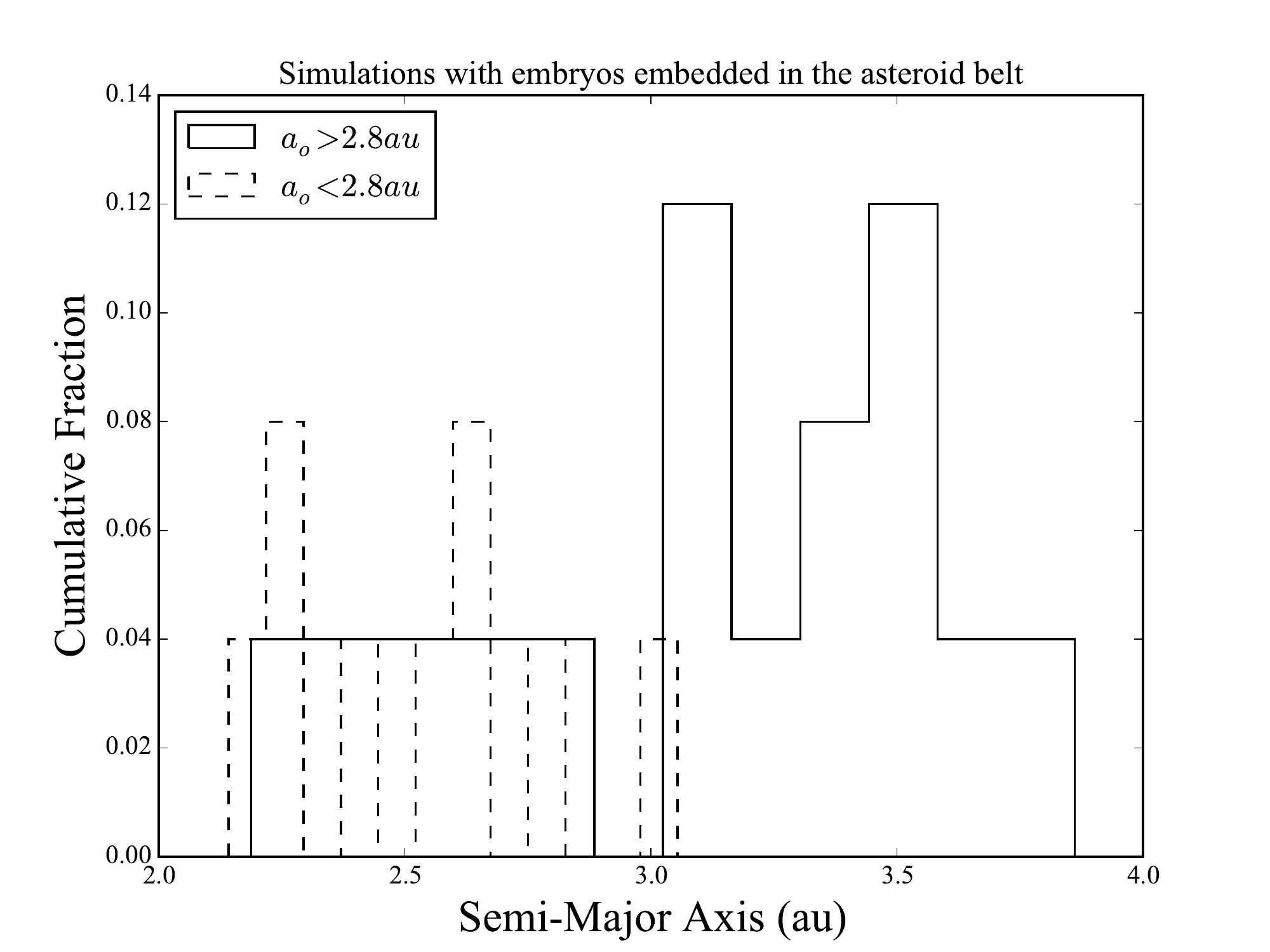}
	\caption{Radial mixing of silicate rich (S-type; dashed line) and carbonaceous (C-type; solid line) bodies in the actual AB (top left panel; all known asteroids with spectral classifications and D$>$50 km) and our various simulation sets.  The different curves in the other three plots (Interact, Test Particle and Embryos) compare the final distributions of all asteroids that began the integration with $a<$2.8 au (dashed lines) to those that started with $a>$2.8 au (solid lines).  Only asteroids not on planet crossing orbits are considered.}
	\label{fig:cs}
\end{figure*}

We adopt a simple model to test how the giant planet instability mixes material from different radial origins in the AB.  We assume that our simulations begin with the outer AB ($a>$2.8 au) highly populated with C-types, and the inner MB dominated by S-types.  In figure \ref{fig:cs} we plot the final distributions of asteroids originating from inside and outside 2.8 au for our three simulation sets.  Though this figure is biased towards simulations with less depletion, our fully interacting runs provide the best match to the current two distributions.  The majority of the total belt material is in the form of C-types, and the interior population of S-types is significantly mixed with the outer C-types.  Specifically, 67$\%$ of all remaining asteroids in these runs are considered C-types (originate outside 2.8 au), as compared to 60$\%$ of all modern asteroids with D$>$50 km.  43$\%$ of those same asteroids that reside in the inner MB (a$<$2.8 au) are C-types (compared with 49$\%$ in the solar system).  Furthermore, our runs with more violent instabilities (and large depletion fractions) where $P_{S}/P_{J}$ finished greater than 2.8 display a similar mixing trend.  In these runs 76$\%$ of all surviving asteroids are C-types.  Thus our mixing results are loosely consistent across the range of our instabilities.  Though this trend is similar for all 3 simulation sets, the test particle runs over-deplete both the inner MB and total S-type inventory.  Contrarily, the Embryo runs scatter a significant number of C-types inward, but fail to scatter the primordial S-types outward. 

In general, the test particle only runs (1a-8a) over-deplete the inner MB of S-types because there is no dynamical friction present to save material with a$<$2.5 au from loss when the $\nu_{16}$ and $\nu_{6}$ secular resonances sweep through the belt.  Indeed, our test particle simulations loose an average of 29$\%$ more mass in the inner MB (2.0$<$a$<$2.5 au) in the 100 Kyr after the first ice giant ejection than in our other simulations.  This is consistent with other studies of resonant sweeping in the AB \citep{walshmorb11}.  However, it is also possible that this is the result of small number statistics and the limited number of simulations used in our study.  When disk self-gravity is included (runs 1-8), the overall distribution is a much better match to that of the actual solar system.

Though the populations of S and C-types in runs 1b and 2b are not as well mixed as those in runs 1-8, the small number statistics make it difficult to say whether the difference is significant.  Indeed, runs 1b and 2b both experience uncharacteristically weak instabilities when compared with our other runs (see figure \ref{fig:e55}).  Both only excite $e_{55}$ to less than half the modern value, and finish with an $AMD_{JS}$ less than a third that of $AMD_{JS,SS}$.  Since the giant planets' resonant perturbations were less strong in these simulations, it makes sense that the final MB populations are less mixed than those of the other simulation sets.

Nevertheless, the radial mixing results of our interaction set are encouraging and in decent agreement with the modern solar system's distribution (though this assumes perfect zoning at the beginning of our runs, and is biased towards runs with less depletion).  Furthermore, since the Kirkwood gaps have yet to fully shape the radial structure of the MB at t$=$200 Myr, we need not expect the shapes (in particular the inter-gap peaks) of the distributions to match exactly \citep{moons95,petit01,obrien07,deienno16,deienno18}.

\subsection{Orbital Structure}

\begin{table}
\centering
\begin{tabular}{c c c c c}
\hline
Run & $AMD_{JS}$ & $P_{S}/P_{J}$ & $M_{AB,f}/M_{AB,fi}$ & $\nu_{6}$ ratio \\
\hline
1  & 0.048 & 2.35 & 0.23 & 0.31 \\
2 & 0.097 & 1.93 & 0.34 & 2.31 \\
3 & 2.33 & 2.77 & 0.0014 & inf \\
4 & 74.3 & 21.5 &  0 & N/A \\
5 & 1.72 & 3.94 & 0.005 & inf \\
6 & 0.057 & 2.08 & 0.030 & 4.25 \\
7 & 47.7 & 17.9 & 0 & N/A \\
8 & 0.093 & 3.94 & 0.17 & 0.50 \\
1a & 1.03 & 5.16 & 0.15 & 0.60 \\
2a & 0.096 & 2.53 & 0.088 & 2.00 \\
3a & 5.84 & 4.12 & 0.0033 & 3.00 \\
4a & N/A & N/A & 0.036 & 0.91 \\
5a & 0.074 & 2.41 & 0.14 & 1.45 \\
6a & 0.78 & 4.37 & 0.0064 & 0.67 \\
7a & 21.29 & 6.19 & 0 & N/A \\
8a & 12.7 & 8.02 & 0.022 & 1.67 \\
1b & 0.31 & 2.70 & 0.13 & 1.00 \\
2b & 0.0086 & 2.23 & 0.18 & inf \\
\hline
\end{tabular}
\caption{Table of results for our 18 runs.  The columns are as follows: (1) the run number, (2) the AMD of the Jupiter/Saturn system, (3) Jupiter and Saturn period ratio, (4) the run's AB depletion fraction, and (5) the ratio of asteroids above to below the $\nu_{6}$ ratio.  Only asteroids not on planet crossing orbits are considered.}
\label{table:nu6}
\end{table}                                                                                      

Most studies of the MB's detailed structure attempt to match a simulated belt's a/e and a/i distributions with that of the actual MB.  The actual MB orbital distribution is well characterized in a/e space by a uniform distribution of excited orbits that inhabit the entire allowed parameter space (that is to say all excited orbits that are not planet crossing).  However, the present day MB's a/i distribution is peculiar in that it is largely devoid of high inclination asteroids above the $\nu_{6}$ secular resonance.  The inability to match the ratio of simulated asteroids in the inner MB above to below this resonance with that of the solar system (often cited as $\sim$0.08; calculated by taking the 100 biggest asteroids with a$<$2.8 au) is a common pitfall of many studies of MB dynamics \citep{walshmorb11,deienno16}.  However, the MB's radial mass distribution profile is far from uniform because half of its mass is concentrated in just 4 asteroids.  From the standpoint of mass distribution, the MB's structure seems to be better characterized by just a handful of massive bodies embedded in a sea of objects several orders of magnitude smaller (many of which are likely the products of collisional grinding; \citet{bottke05a,bottke15,dermott18}).  Indeed, the $\nu_{6}$ ratio calculated for all known asteroids with a$<$2.8 au with respect to number of asteroids (0.04), is a significantly smaller than when it is calculated with respect to mass (0.14).  The reason for this is that the third largest asteroid, Pallas, has an inclination above the $\nu_{6}$ resonance (33$^{\circ}$).

In the sense that the AB's structure is dominated by a small number of asteroids, the third largest of which is above the $\nu_{6}$ resonance, our simulations are largely successful.  Since our initial asteroids have masses within an order of magnitude or so of the actual larger asteroids in the MB, it is encouraging that the majority of our simulations deplete the MB down to just a few of such objects.  Furthermore, the MB structure is also dominated by multiple large collisional families that formed after the epoch of terrestrial planet formation \citep{bottke15,dermott18}.  Thus it is supremely difficult to correlate the current belt's orbital and size distributions with those at t$=$200 Myr \citep{bottke05a}.

\begin{figure}
	\centering
	\includegraphics[width=.5\textwidth]{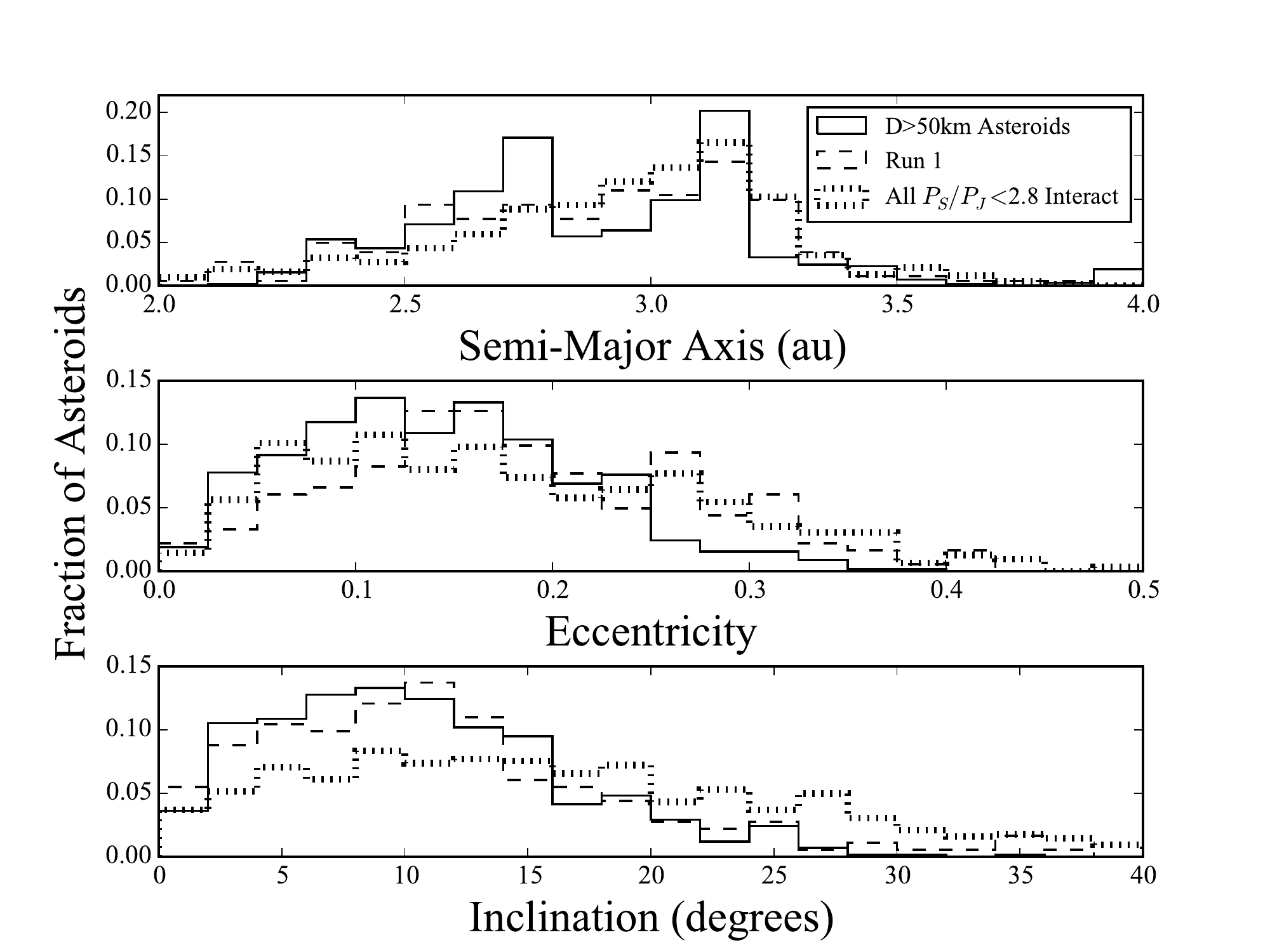}
	\caption{Histograms of semi-major axis (top panel), eccentricity (middle panel), and inclination (bottom panel) distributions  for the modern asteroid belt (solid lines; H $<$ 9.7, D $>$ 50 km), run 1 (dashed lines) and all interaction systems with $P_{S}/P_{J}<$ 2.8 (dot-dashed lines).  Only asteroids not on planet crossing orbits are considered.}
	\label{fig:aei}
\end{figure}
 
 \begin{figure}
	\centering
	\includegraphics[width=.5\textwidth]{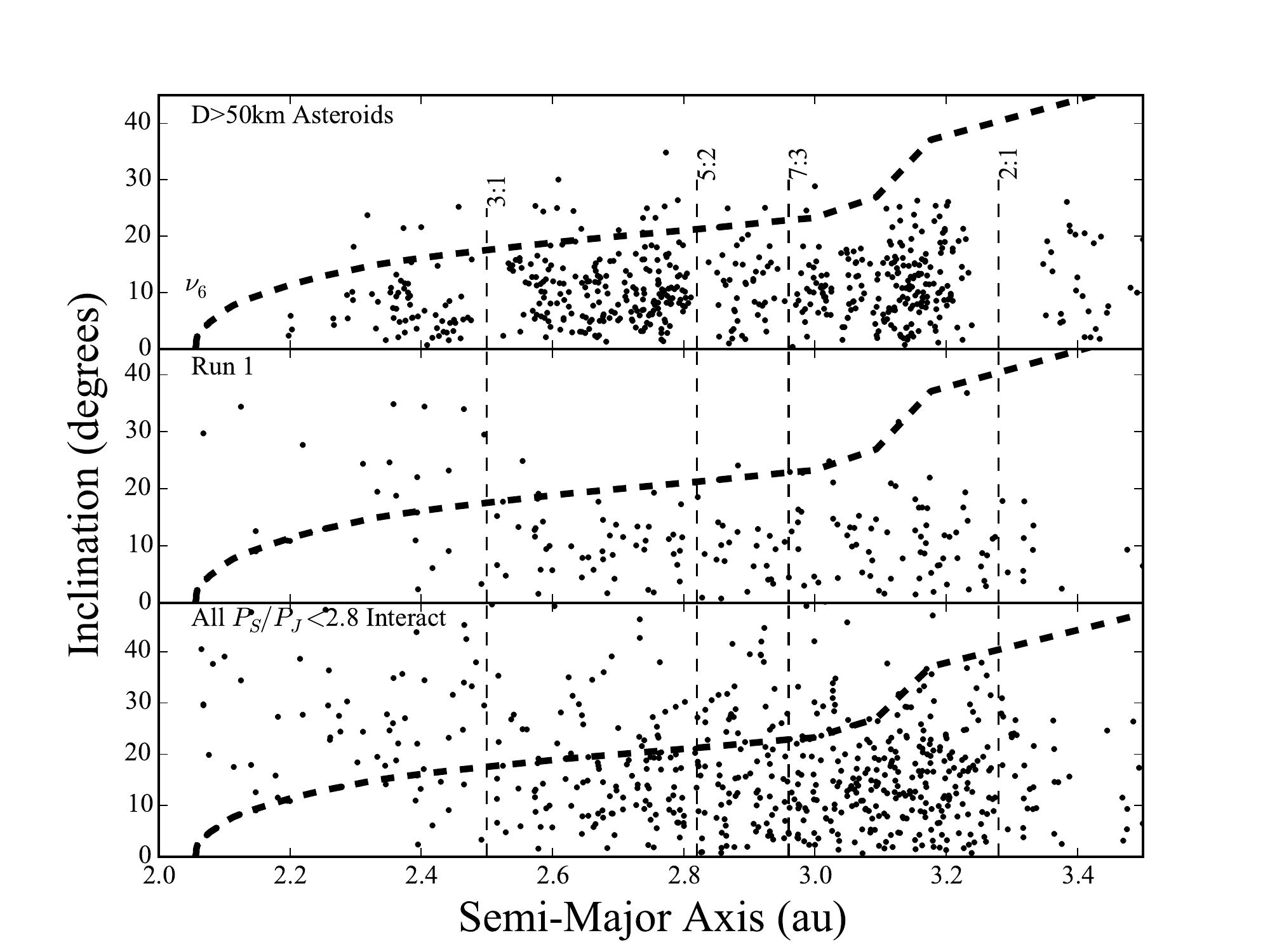}
	\caption{Semi-major axis/inclination plot for the modern asteroid belt (top panel; H $<$ 9.7, D $>$ 50 km), run 1, and all interaction systems with $P_{S}/P_{J}<$ 2.8.  Only asteroids not on planet crossing orbits are considered.  The vertical dashed lines represent the locations of the important mean motion resonances with Jupiter.  The bold dashed lines indicate the current location of the $\nu_{6}$ secular resonance.}
	\label{fig:inc}
\end{figure}

 \begin{figure}
	\centering
	\includegraphics[width=.5\textwidth]{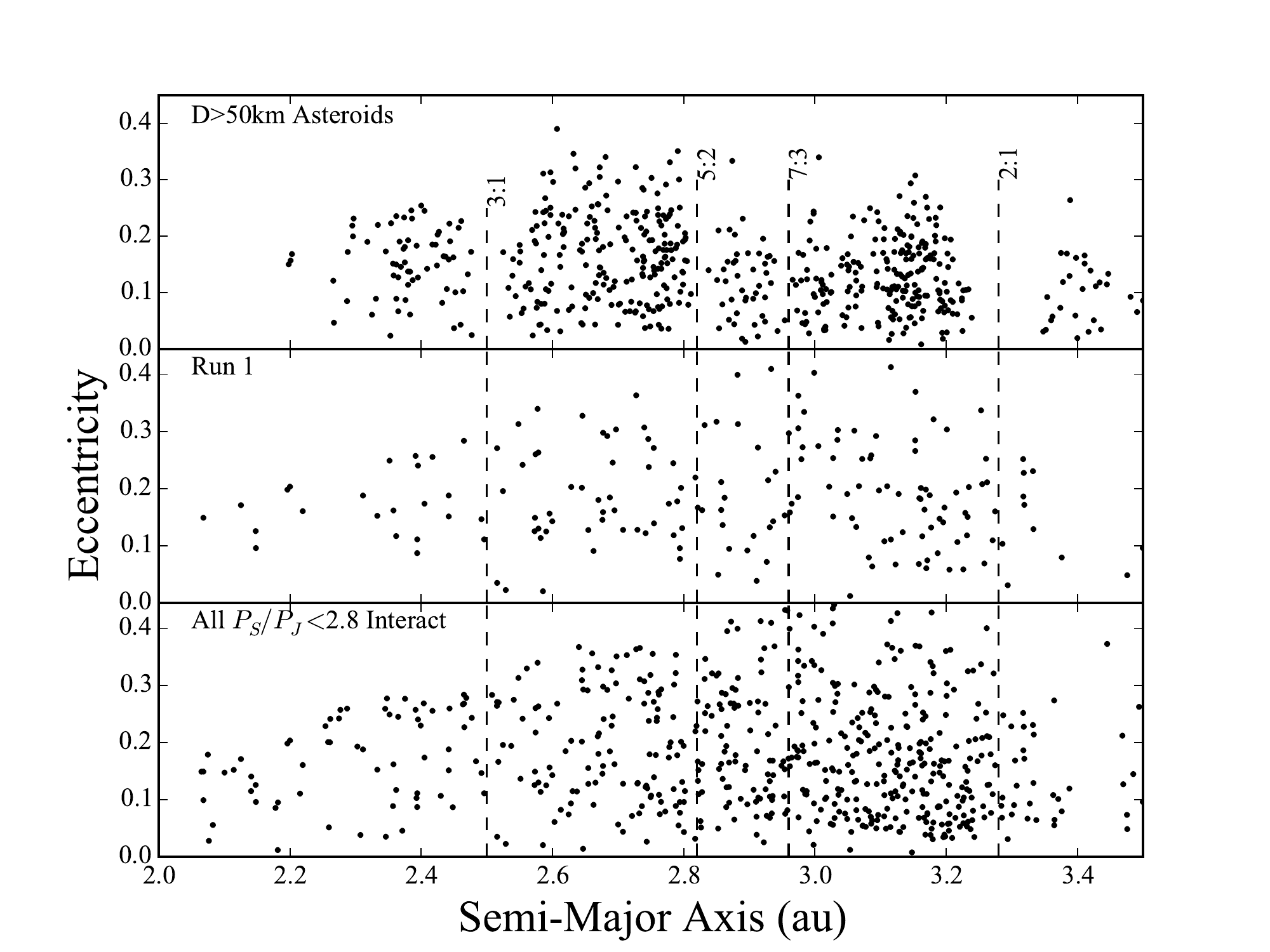}
	\caption{Semi-major axis/eccentricity plot for the modern asteroid belt (top panel; H $<$ 9.7, D $>$ 50 km), run 1, and all interaction systems with $P_{S}/P_{J}<$ 2.8.  Only asteroids not on planet crossing orbits are considered.  The vertical dashed lines represent the locations of the important mean motion resonances with Jupiter.}
	\label{fig:ecc}
\end{figure}

Nevertheless, our simulated belt's eccentricity and inclination distributions are fairly similar to the real distribution.  In figures \ref{fig:aei}, \ref{fig:inc} and \ref{fig:ecc}, we compare the orbital distributions for all bright (H$<$9.7) asteroids larger than 50 km with our results from run 1, and all of our interaction runs where Jupiter and Saturn's period ratio finished below 2.8.  We select run 1 in this figure because it finished with the most MB objects of any interaction run with $P_{S}/P_{J}<$ 2.8 and $N_{GP}=$ 4.  Thus we provide an example of a single successful distribution without co-adding results.  However, this plot should be taken in context with the fact that run 1 finishes with a low $AMD_{JS,SS}$, and significantly under-excited $e_{55}$ and $e_{56}$ amplitudes.  Furthermore, the co-added plot is biased toward similar simulations with lower depletion.

In keeping with previous works, our simulations consistently struggle to replicate the modern AB's structure about the $\nu_{6}$ secular resonance.  \citet{walshmorb11} reported a ratio of $\sim$5.2 in a smooth migration scenario.  \citet{deienno16} evaluated a post-Grand Tack version of the late Nice Model and determined a ratio $\sim$1.2 (0.10 when removing post-Grand Tack asteroids with $i>$20$^\circ$).  Studies of an early instability have shown promise in replicating some aspects of the belt's structure, but still fall short when it comes to the $\nu_{6}$ ratio.  \citet{clement18} co-added the remaining asteroids from successful simulations ($P_{S}/P_{J}<$2.8) and found the ratio to be $\sim$0.73, but the value was much worse when only considering the inner MB asteroids with $a<$2.5 au ($\sim$2.2).  In a similar study, \citet{deienno18} considered a ``Jumping Jupiter'' style instability and found the ratio to be $\sim$1.3 (0.07 when removing post-instability asteroids with $i>$20$^\circ$).  In general, our simulations (table \ref{table:nu6}) finish with ratios of similar order to those found in previous early instability studies.  Our best ratio occurred in run 1 (0.31).  Because the sculpting mechanisms in each instability are unique, it is difficult to extrapolate a correlation between depletion and the final $\nu_{6}$ ratio.

Two of our runs finished with the entire inner MB population's inclinations above the $\nu_{6}$ resonance.  This occurs when the location of the $\nu_{6}$ resonance drags more smoothly through the inner MB.  Instabilities where the migration of the giant planet's orbits and respective resonances is more chaotic are less likely to excite the entire region uniformly, and finish with better ratios.  This is also the case in the nominal ``Jumping Jupiter'' style instability \citep{bras09,nesvorny13,deienno18}.  In either scenario, the location of the resonance does not spend a significant amount of time in any one location during the planetary instability.  

In spite of several moderately successful runs (eg: runs 1, 8, 1a, 4a and 6a) that finish with $\nu_{6}$ ratios less than unity, reproducing the inner MB's a/i structure is still an outstanding problem for the early instability model.  However, the MB's orbital distribution has evolved as the result of family-forming events over the past $\sim$4 Gyr \citep{bottke15}.  In fact, \citet{dermott18} estimated that $\sim$85$\%$ of inner MB asteroids originated from one of just five original collisional families (Flora, Vesta, Nysa, Polana and Eulalia).  Thus, it is possible that the primordial AB contained very few objects.  If Pallas was one of this primordial population of a few, ``large'' asteroids, and its high inclination (above the $\nu_{6}$ resonance) is primordial, our results may not be particularly problematic.  Indeed, planet formation simulations that include collisional fragmentation \citep{clement18_frag} indicate that collisional fragments tend to preferentially populate the inclination parameter space below the $\nu_{6}$ secular resonance.  Thus, future work on the topic must utilize similar GPU accelerated integration schemes to study how collisional fragmentation shapes the young asteroid belt.  If it turns out that the fragmentation process dramatically alters the primordial belt's size distribution, it is possible that the dynamical friction generated by a diverse population of fragments could save material from loss during the instability.  If that were the case, the depletion values discussed in section 3.1 might be overestimated.  

\subsection{Seeded Embryos}

We implanted 4 Mercury massed embryos in the primordial MB in two of our simulations (runs 1b and 2b).  As discussed in the introduction, early dynamical studies of terrestrial planet formation often argued that primordial Moon to Mars massed planet embryos might explain the AB's substantial early mass loss \citep{wetherill92,chamb_weth01,chambers01,petit01}.  To prevent additional semi-major axis gaps in the belt's structure from being fossilized, embryos larger than Mercury could not have survived the planet formation process \citep{chambers07,ray09a,obrien11}.  Thus the problem with the depletion via primordial embryos model is two-fold.  First, recent studies of pebble accretion during the gas disk phase indicate that large embryos may not have formed so far out in the terrestrial disk \citep{levison15}.  Second, studies that include large embryos throughout the entire MB region often fail to destabilize all of them within the $\sim$200 Myr of terrestrial planet formation \citep{chambers01,ray09a}.  Indeed only 54$\%$ of the best set of the early instability simulations in \citet{clement18} deplete all primordial embryos from the AB.  

Our simulations that began with four, Mercury-massed planet embryos in the MB provide poor matches to the current solar system for several reasons.  Both simulations  finish with a $\sim$0.25 $M_{\oplus}$ planet in the MB on a stable orbit.  In contrast, only one of our other 16 systems finishes with an object larger than Mercury in the MB region.  In those runs, accretion began slowly in the belt during the initial 1 Myr of integration that did not include any giant planet evolution (\citet{koko_ida_96}; see discussion in section 2).  This growth was then aborted when the Nice Model instability ensued \citep{clement18}.  While the instabilities in both runs were relatively weak in terms of the excitation of the giant planets ($AMD_{JS}$, $e_{55}$ and $e_{56}$), neither system experienced substantially more depletion than did other runs with similar giant planet outcomes (figure \ref{fig:e55}).  Thus, our results indicate that excitation from the giant planets, rather that from embedded embryos, is the dominant depletion mechanism in the early instability scenario \citep{clement18}.  Though our current study's sample of just 2 embryo systems is small, our other successful simulations consistently depleted a substantial percentage of AB material ($\sim$99-99.9$\%$).  Therefore, embedding embryos in the primordial MB population is not necessary within the context of the early instability scenario to deplete the MB by $\sim$3 orders of magnitude.

\subsection{The formation of Veneneia}

As the instability excites asteroids on to orbits where they are ejected from the solar system, many continue to cross the MB region for some time before being lost completely.  In our fully self-interacting runs, these excited objects can still undergo collisions with stable MB asteroids.  When we evaluate the accretion histories of asteroids in our 4 interaction runs that finished with $P_{S}/P_{J}<$2.8, we find that the results are highly dependent on the instability's strength.  Runs 1 and 2 each experienced weaker instabilities, and finished with more asteroids, and low values of $AMD_{JS}$, $e_{55}$ and $e_{56}$.  Asteroids in these instabilities undergo an average of $\sim$0.7 collisions over the 200 Myr simulation duration.  On the other hand, runs 3 and 6 experienced more violent instabilities. In those runs, surviving asteroids experienced no accretion events.  Since our simulations only model larger asteroids (similar to Ceres and Vesta), we can  loosely correlate the accretion histories of our simulated MB particles to that of Vesta.  Because the projectiles that formed the Rheasilvia and Veneneia craters are estimated to have had diameters between $\sim$60-70 km \citep{jutzi13}, we can use the AB's modern size distribution to extrapolate how likely such collisions would be in our models.  The present day belt contains 86 such asteroids.  Therefore, we can conclude that Rheasilvia and Veneneia-forming events would have been too common in runs 1 and 2.  Unfortunately, such a calculation is beyond the resolution of our simulations in runs 3 and 6.  These results, however, are highly biased by the stochastic nature of the instability and the small number of simulations presented in our study.

\section{Conclusions}

In this paper we presented 18 N-body simulations of the effect of an early Nice Model instability \citep{Tsi05,clement18} on a 3000 particle AB.  In 10 of our simulations, the entire AB is modeled as fully self-gravitating. Our simulations show that the early instability scenario can deplete a $\sim$2 $M_{\oplus}$ primordial AB (commensurate with the minimum mass solar nebula; \citet{mmsn}) at the $\sim$99-99.9$\%$ level.  The level of depletion experienced by a particular system is related to the unique evolution of the giant planets within the chaotic orbital instability.  Since each instability is unique, and it is impossible to know the exact evolution of the giant planets in the young solar system's own instability, our work establishes a spectrum outcomes on which we place the solar system.  In particular, depletion increases with higher values of $AMD_{JS}$ and $e_{55}$ (the amplitude of Jupiter's $g_{5}$ eigenmode).  Systems that finish with values of $AMD_{JS}$ and $e_{55}$ similar to those of the present solar system tend to deplete the AB by two to three orders of magnitude.

Assuming the outer primordial AB was populated with volatile rich C-type asteroids (either because of a primordial ``snow-line'' or via scattering during the giant planet growth phase as described in \citet{ray17}), our fully self-gravitating simulations are very successful at reproducing the observed distribution of S and C-types.  Contrarily, radial mixing is not as strong in our simulations that only consider asteroids as test particles.  Furthermore, these test particle only simulations typically over-erode the inner MB region (2.0$<$a$<$2.5 au).  Dynamical friction between asteroids helps to de-excite orbits in the region, and prevent this erosion in our fully self-interacting systems.

In general, our systems adequately replicate the eccentricity and inclination distributions of the modern AB.  However, consistent with various previous studies \citep{walshmorb11,deienno16,clement18,deienno18}, our simulations often fail to replicate the AB's population about the $\nu_{6}$ secular resonance.  However, the MB size distribution has evolved significantly since the epoch of terrestrial planet formation \citep{bottke05a}, and its orbital distribution has likely been altered via the formation of collisional families \citep{dermott18}.  Since the MB's mass profile is dominated by just a few asteroids, one of which (Pallas) is on a highly inclined orbit above the $\nu_{6}$ resonance, our systems still broadly replicate the distribution of the most massive asteroids.  
 
 Our results should be taken in the appropriate context given the sensitivity of the AB's depletion and final structure to the particular dynamics of the giant planet instability.  It is impossible to know the exact evolutionary path followed by the giant planets during the Nice Model instability.  Furthermore, none of our final giant planet systems are exact matches to the modern solar system.  Therefore, our study can only correlate specific dynamical qualities of the present solar system with those of our simulations.  Nevertheless, our fully self-gravitating ABs are largely successful at replicating the modern belt for several reasons.  These include a number of systems with low ratios of asteroids above to below the $\nu_{6}$ resonance, accurate radial distributions of S and C-types, and relatively high levels of MB depletion.  
 
 While our work demonstrates that GPU accelerated N-body simulations offer an unprecedented opportunity to explore higher-N systems, each of our simulations still requires large amounts of computing resources. Specifically, each simulation in this paper required the same number of node hours to complete as 32 conventional (CPU only) integrations of terrestrial planet formation using standard initial conditions \citep{chambers01,clement18}.  In spite of generating a suite of simulated asteroid belts sculpted by giant planet instabilities, the stochasticity of instabilities and the small number of surviving asteroids prevents us from developing detailed predictions of asteroid belt structure as a function of instability characteristics. Doing so will likely require still more extensive sets of instability simulations.
 
\section*{Acknowledgments}

This material is based upon research supported by the Chateaubriand Fellowship of the Office for Science and Technology of the Embassy of France in the United States.  We thank Rogerio Deienno and Bill Bottke for thoughtful comments and insight.  M.S.C. and N.A.K. thank the National Science Foundation for support under award AST-1615975.  S.N.R. thanks the Agence Nationale pour la Recherche for support via grant ANR-13-BS05-0003-002 (grant MOJO) and acknowledges NASA Astrobiology Institute’s Virtual Planetary Laboratory Lead Team, funded via the NASA Astrobiology Institute under solicitation NNH12ZDA002C and cooperative agreement no. NNA13AA93A.  This research is part of the Blue Waters sustained-petascale computing project, which is supported by the National Science Foundation (awards OCI-0725070 and ACI-1238993) and the state of Illinois. Blue Waters is a joint effort of the University of Illinois at Urbana-Champaign and its National Center for Supercomputing Applications \citep{bw1,bw2}.

\bibliographystyle{apj}
\newcommand{\sci}{$Science$ }
\bibliography{ab.bib}
\end{document}